\documentclass[aps,prd,superscriptaddress,twoside,twocolumn,nofootinbib,10pt,%
showpacs]{revtex4-1}

\usepackage{amsmath,amssymb}
\usepackage{multirow}
\usepackage{dcolumn}
\usepackage{ulem} 
\usepackage[usenames]{color}

\allowdisplaybreaks

\usepackage{multirow}
\begin{document}


\title{Degeneracy of doubly heavy baryons from heavy quark symmetry}

\author{Yong-Liang Ma}
\email{yongliangma@jlu.edu.cn}
\affiliation{Center of Theoretical Physics and College of Physics, Jilin University, Changchun,
130012, China}

\author{Masayasu Harada}
\email{harada@hken.phys.nagoya-u.ac.jp}
\affiliation{Department of Physics,  Nagoya University, Nagoya, 464-8602, Japan}

\date{\today}
\begin{abstract}
The spectroscopy of the doubly heavy baryons including different heavy quarks is studied based on
the heavy quark symmetry of QCD. We point out that, when the two heavy quarks are in $S$-wave, these baryons with a certain spin $j_l$ of the light cloud can be classified into two sets: a heavy quark singlet with total spin of $j=j_l$ and a heavy quark multiplet with $j= (j_l+1) , j_l ,\ldots \vert j_l-1\vert$, all the baryons in these two sets have the same mass and, the baryons with the same quantum numbers in these two sets do not mix with each other. We finally point out that the strong decay of the first excited baryon with light spin $j_l = 1/2$ to the ground state and one-pion is determined by the mass splitting through the generalized Goldberger--Treiman relation.

\end{abstract}
\pacs{
14.20.-c,12.39.Hg,11.30.Rd,13.30.-a
}

\maketitle

The physics of heavy hadrons has become a hot subject in particle and nuclear physics because of the observations of a large amount of such states during the last decade in scientific facilities. With the collection of more data and updating of facilities,
more and more states must be observed in the present and upcoming facilities such as BESIII, LHCb and Belle II. The observation of the hidden charm pentaquark state at LHCb~\cite{Aaij:2015tga} strongly indicates that it is the time to study baryons with two heavy quarks, the doubly heavy baryons (DHBs).

The DHB is an immediate prediction from the quark model. Even though the DHBs have been extensively discussed theoretically using several models in the literature~\cite{Kiselev:2001fw,Likhoded:2009zz,Savage:1990di,Gershtein:2000nx,Kiselev:1999zj,Kiselev:2000jb,Kiselev:2002iy,Vijande:2004at,Ebert:2004ck,Brambilla:2005yk,Hu:2005gf,Zhang:2008rt,Tang:2011fv,Brodsky:2011zs,Karliner:2014gca,Brown:2014ena,Sun:2014aya,Ma:2015lba}, there are controversial results in the experimental hunting~\cite{Agashe:2014kda,Mattson:2002vu,Ocherashvili:2004hi,Ratti:2003ez,Chistov:2006zj,Aubert:2006qw,Aaij:2013voa,Kato:2013ynr,Agashe:2014kda,Mattson:2002vu,Ocherashvili:2004hi}. It is useful to provide some more theoretical guidance by different models for the future experimental search.

In this work, we will focus on the spectroscopy of the DHBs with different heavy quarks based on the heavy quark symmetry (see, e.g. Ref.~\cite{Manohar:2000dt} for a review).
In a DHB with different heavy quarks, say $Q$ and $Q^\prime$, the two heavy quarks in the $S$-wave can form a spin singlet and a spin triplet. For simplicity, we write the spin singlet and the spin triplet as $\bar{\Phi}^{(QQ^\prime)}$ and $\bar{\Phi}_\mu^{(QQ^\prime)}$, respectively, with both $\bar{\Phi}^{(QQ')}$ and $\bar{\Phi}_\mu^{(QQ')}$ being the color anti-triplet. Due to the spin-flavor symmetry in the heavy quark limit, $\bar{\Phi}^{(QQ')}$ and $\bar{\Phi}_\mu^{(QQ'}$ have the same mass, i.e.,
\begin{equation}
M(\bar{\Phi}^{(QQ')}) = M(\bar{\Phi}_\mu^{(QQ')}) \ .
\end{equation}

The interaction between the heavy diquarks, $\Phi^{(QQ^\prime)}$ and $\bar{\Phi}_\mu^{(QQ^\prime)}$, and gluons can be easily written down by considering that both $\Phi^{(QQ^\prime)}$ and $\bar{\Phi}_\mu^{(QQ^\prime)}$ are the color anti-triplets. By taking the heavy quark limit, the effective Lagrangian for the diquarks is expressed as
\begin{eqnarray}
\mathcal{L}_{\rm eff}^{\Phi}& = & \bar{\Phi}^{(QQ^\prime)} i\,v_\nu (\partial^\nu + i g G^\nu ) \bar{\Phi}^{(QQ^\prime)\dag} \nonumber\\
& & {} +
\bar{\Phi}^{(QQ^\prime)\mu} i\,v_\nu (\partial^\nu + i g G^\nu ) \bar{\Phi}^{(QQ^\prime)\dag}_\mu \ ,\label{eq:LagPhiG}
\end{eqnarray}
where $v_\nu$ is the velocity of the diquarks, $G^\nu$ is the gluon field and $g$ is the gauge coupling constant of QCD.
Effective Lagrangian \eqref{eq:LagPhiG} implies that the two diquarks have the same interaction with the gluon which combines them to a light degree of freedom (``Brown muck'') to form two types of heavy baryons.

Now, we are in the position to study the mass relation of the DHBs with quark content $QQ^\prime q$ where $q$ stands for a light quark constituent.

We first consider the DHBs in the ground state. In such a case, we schematically write
the DHBs as $D_\textbf{Q} \equiv \bar{\Phi}^{(QQ^\prime)} q$ and $D_\textbf{Q}^\mu \equiv \bar{\Phi}^{(QQ^\prime)\mu} q$, where $q$ symbolically denotes the Brown muck in the ground state which carries the spin-parity $j^P_l = \frac{1}{2}^+$.
Since the scalar diquark $\bar{\Phi}^{(QQ')}$ carries spin zero, the spin-parity of the $D_\textbf{Q}$ is $j^P = \frac{1}{2}^+$. Therefore $D_\textbf{Q}$ is a heavy quark singlet. On the other hand, since the axial-vector diquark $\bar{\Phi}_\mu^{(QQ^\prime)}$ carries spin one, the spin-parity of the $D_\textbf{Q}^\mu$  is either
$j^P = \frac{1}{2}^+$ or $\frac{3}{2}^+$ which forms a heavy quark doublet. With respect to Eq.~\eqref{eq:LagPhiG}, we see that
the singlet $D_\textbf{Q}$ and doublet $D_\textbf{Q}^\mu$ have the same mass, i.e.,
\begin{equation}
M(D_\textbf{Q}) = M(D_\textbf{Q}^\mu) \ .
\end{equation}
Furthermore, the states with the same quantum number, explicitly those with $j^P = \frac{1}{2}^+$, in $D_\textbf{Q}$ and $D_\textbf{Q}^\mu$ cannot mix due to the difficulty of the heavy quark spin flipping. Then, we arrive at the conclusion that, in the heavy quark limit, the ground states of the DHBs with different heavy quarks form a heavy quark singlet and a heavy quark doublet which are classified by the total spin of the heavy diquark included in them and the DHBs in these two sets have the same mass.

Next, we consider DHBs with the first orbital excitation with relative angular momentum between the light quark and heavy diquark source $l = 1$. In such a case, the quantum numbers of the Brown muck could be $j_l^P = \frac{1}{2}^-$ and $\frac{3}{2}^-$. Combining the $j_l^P = \frac{1}{2}^-$ light component with the heavy component $\bar{\Phi}^{(QQ')}$ and $\bar{\Phi}_\mu^{(QQ')}$
one can form a heavy quark singlet $N_\textbf{Q}$ with $j^P=\frac{1}{2}^- $ and a heavy quark doublet $N_\textbf{Q}^\mu$ with $j^P = \frac{1}{2}^-$ and $\frac{3}{2}^-$, respectively.
In analogy to the discussion made to the ground states, the baryons in the heavy quark singlet $N_\textbf{Q}$ and
those in the heavy quark triplet $N_\textbf{Q}^\mu$ have the same mass. When we combine the $j_l^P = \frac{3}{2}^-$ component to the heavy diquarks one gets a heavy quark singlet $T_\textbf{Q}^\mu$ with quantum numbers $j^P = \frac{3}{2}^-$ and a heavy quark triplet $T_\textbf{Q}^{\prime\mu}$ with quantum numbers $j^P = \frac{5}{2}^-,\frac{3}{2}^-,\frac{1}{2}^-$. Again, from the effective Lagrangian~\eqref{eq:LagPhiG} we conclude that all the baryons in the heavy quark singlet $T_\textbf{Q}^{\mu}$ and heavy quark triplet $T_\textbf{Q}^{\prime\mu}$ have the same mass and the two baryons with quantum numbers $j^P = \frac{3}{2}^-$ decouple from each other.

From the above discussion, we arrive at our conclusion that, {\it for DHBs with total light spin-parity $j_l^P$, they can be classified into two sets: a heavy quark singlet with quantum numbers $j^P = j_l^P$ and a heavy quark multiplet with quantum numbers $j^P = (j_l+1)^P, \cdots, (|j_l -1|)^P$. These baryons have degenerate masses and the two baryons with quantum numbers $j^P = j_l^P$ in these two sets do not mix to each other due to heavy quark spin conservation in the heavy quark limit}. Examples of the ground states and first excited states are summarized in Table.~\ref{tab:Deg}.

\begin{table}[htb]
\begin{tabular}{*{10}{r}}
\hline\hline
\quad $J_Q$ &\quad $l$ &\qquad $j_l^P$ \;\; &\quad $j^P$ \;\;\;\;\;\; & Mass relation \\
\hline
$0$ \, & $0$  & $\frac{1}{2}^+$ \;\; & $\frac{1}{2}^+$ \;\;\;\;\;\; & \multirow{2}*{degenerate}\;\;\\
$1$ \, & $0$  & $\frac{1}{2}^+$ \;\; &\quad $(\frac{3}{2}^+,\frac{1}{2}^+)$\;\;\;\; \\
\hline
$0$ \, & $1$  & $\frac{1}{2}^-$ \;\; & $\frac{1}{2}^-$ \;\;\;\;\;\; & \multirow{2}*{degenerate}\;\;\\
$1$ \, & $1$  & $\frac{1}{2}^-$ \;\; & $(\frac{3}{2}^-,\frac{1}{2}^-)$\;\;\;\; \\
\hline
$0$ \, & $1$  & $\frac{3}{2}^-$ \;\; & $\frac{3}{2}^-$ \;\;\;\;\;\;\; & \multirow{2}*{degenerate}\;\;\\
$1$ \, & $1$  & $\frac{3}{2}^-$ \;\; & $(\frac{5}{2}^-,\frac{3}{2}^-,\frac{1}{2}^-)$ \\
\hline\hline
\end{tabular}
\caption{\label{tab:Deg} Pattern of mass degeneracy of the ground states and the first excited states. Notations are explained in the main text.
}
\end{table}

After the above discussion on the pattern of the mass degeneracy, we turn to the strong decays of the DHBs. Because the heavy quarks in a DHB have large masses, the light quark in the DHB sees the heavy diquark as a local source of gluon, which makes the picture of the DHB analogues to the heavy-light meson~\cite{Savage:1990di}. Then, to analyze the strong decays of the DHBs with $j_l^P=\frac{1}{2}^-$, we use the chiral partner structure applied in the heavy-light meson sector~\cite{Nowak:1992um,Bardeen:1993ae}. There, the heavy-quark doublet including $j^P=0^-$ and $j^P=1^-$ heavy-light mesons is regarded as the chiral partner of the doublet including $j^P=0^+$ and $j^P=1^+$ heavy-light mesons. The coupling strengths of interactions between two doublets are determined from the mass differences through generalized Goldberger-Treiman relations, which are in good agreement with experiments~\cite{Bardeen:2003kt,Nowak:2003ra,Nowak:2004jg}. The difference here is that, in the heavy-light meson sector the heavy component is a heavy quark but in the DHB sector the heavy component is a heavy diquark made of two heavy quarks. But this difference does not affect the chiral structure which is controlled by the light quark degree of freedom in a hadron. In addition, since it takes much more energy to excite the heavy diquark constituent in a DHB, we regard the excited DHB as those with the light quark excitation in it.
Similar to the heavy-light meson sector, we regard the DHBs with the Brown muck of $j_l^P = \frac{1}{2}^-$, i.e., the heavy quark singlet $N_\textbf{Q}$ and heavy quark doublet $N_\textbf{Q}^\mu$, as the chiral partners to the ground states, i.e., $D_\textbf{Q}$ and $D_\textbf{Q}^\mu$, respectively. Note that, as in the heavy-light meson sector~\cite{Nowak:1993vc}, the extension of the present discussion to the excited states is straightforward.

To accommodate the chiral dynamics in the DHB sector, along Ref.~\cite{Ma:2015lba}, we introduce the left- and right-handed DHB fields $D_{\textbf{Q};L,R}^{(\mu)}$ which, at the quark level, are schematically written as $D_{\textbf{Q};L,R}^{(\mu)} \sim \bar{\bm{\Phi}}^{(\mu)} q_{L, R}$. Under chiral transformation, they transform as
\begin{eqnarray}
D_{\textbf{Q};L,R}^{(\mu)} & \to & g_{L,R} D_{\textbf{Q};L,R}^{(\mu)} ,
\label{eq:chiralDLR}
\end{eqnarray}
where
$g_{L,R} \in SU(2)_{L,R}$ when we consider only the up and down quarks in the DHBs.
In terms of $D_{\textbf{Q}}^{(\mu)}$ and $N_{\textbf{Q}}^{(\mu)}$, one can write
\begin{eqnarray}
D_{\textbf{Q};L}^{(\mu)} & = & \frac{1}{\sqrt{2}}\left(D_{\textbf{Q}}^{(\mu)} - i N_{\textbf{Q}}^{(\mu)} \right)
, \nonumber\\
D_{\textbf{Q};R}^{(\mu)} & = & \frac{1}{\sqrt{2}}\left(D_{\textbf{Q}}^{(\mu)} + i N_{\textbf{Q}}^{(\mu)} \right),
\label{eq:ChiralPhys}
\end{eqnarray}
which transform as $D_{\textbf{Q};L, R}^{(\mu)} \leftrightarrow \gamma_0 D_{\textbf{Q};{(\mu)};R, L}$ under
parity transformation and satisfy $v\hspace{-0.17cm}\slash D_{\textbf{Q};L,R}^{(\mu)} =
D_{\textbf{Q};L,R}^{(\mu)}$ and $v_\mu D_{\textbf{Q};L,R}^{\mu} = 0$ for preserving the heavy quark
symmetry and keeping the transversality. Further, for later convenience, we write the DHB doublets $D_\textbf{Q}^{(\mu)}$ and $N_\textbf{Q}^{(\mu)}$ in terms
of the physical states. For $D_\textbf{Q}$ and $N_\textbf{Q}$ we have
\begin{eqnarray}
D_{\textbf{Q}} & = & \frac{1 + v \hspace{-0.17cm}\slash}{2}\Psi_{QQ^\prime}^\prime , \quad N_{\textbf{Q}} = \frac{1 + v \hspace{-0.17cm}\slash}{2}\Psi_{QQ^\prime}^{\prime\ast},
\label{eq:DNPsi}
\end{eqnarray}
where $\Psi'_{QQ'}$ and $\Psi^{\prime\ast}_{QQ'}$ are Dirac spinors for the DHBs with $j^P=\frac{1}{2}^{+}$ and $\frac{1}{2}^-$, respectively.
Note that the parity transformations of $\Psi'_{QQ'}$ and $\Psi^{\prime\ast}_{QQ'}$ are given as
\begin{eqnarray}
{\rm P} &:&~~ \Psi_{QQ^\prime}^{\prime} \to \gamma_0 \Psi_{QQ^\prime}^{\prime} ,\quad \Psi_{QQ^\prime}^{\prime\ast} \to {} -
\gamma_0\Psi_{QQ^\prime}^{\prime\ast}.
\end{eqnarray}
The expressions of $D_\textbf{Q}^\mu$ and $N_\textbf{Q}^\mu$ in terms of physical states and their transformation are given in Ref.~\cite{Ma:2015lba}. We will not repeat them here. In terms of the naming scheme in PDG, $\Psi_{QQ^\prime}^{(\prime)}$ stands for $\Xi_{bc}^{(\prime)}$ and $\Omega_{bc}^{(\prime)}$ for the DHB including un-flavored quark and strange quark, respectively.

Following the procedure used in Ref.~\cite{Ma:2015lba} one can easily construct an effective Lagrangian of $D_\textbf{Q}^{(\mu)}$ and $N_\textbf{Q}^{(\mu)}$ first in the chiral basis in a chiral invariant way and then rewrite it in terms of $D_\textbf{Q}^{(\mu)}$ and $N_\textbf{Q}^{(\mu)}$. There is no coupling between the heavy-quark doublet and heavy quark singlet because of the heavy quark spin conservation. The chiral effective Lagrangian is simply a duplicate of the one given in Ref.~\cite{Ma:2015lba} which is written in the chiral basis as
\begin{widetext}
\begin{eqnarray}
{\cal L}_{\rm B} & = & \bar{D}_{\textbf{Q};L}^{(\mu)} i v\cdot \partial D_{\textbf{Q};(\mu);L} +
\bar{D}_{\textbf{Q};R}^{(\mu)} i v\cdot \partial D_{\textbf{Q};(\mu);R} - \Delta\left( \bar{D}_{\textbf{Q};L}^{(\mu)} D_{\textbf{Q};(\mu);L} + \bar{D}_{\textbf{Q};R}^{(\mu)}
D_{\textbf{Q};(\mu);R} \right) \nonumber\\
& &{} - \frac{1}{2}g_\pi\left( \bar{D}_{\textbf{Q};L}^{(\mu)} M D_{\textbf{Q};(\mu);R} +
\bar{D}_{\textbf{Q};R}^{(\mu)} M^\dagger D_{\textbf{Q};(\mu);L} \right)  + \frac{ig_A}{f_\pi}\left[ \bar{D}_{\textbf{Q};L}^{(\mu)}\gamma_5\gamma^\nu \partial_\nu M
D_{\textbf{Q};(\mu);R}  + \bar{D}_{\textbf{Q};R}^{((\mu))}\gamma_5\gamma^\nu \partial_\nu M^\dagger
D_{\textbf{Q};(\mu);L}\right] ,\nonumber\\
\label{eq:EffecL}
\end{eqnarray}
where $M$ is the light meson field which transforms as $M \to g_L M g_R^\dag$ under chiral transformation.
\end{widetext}
In terms of the scalar and pseudoscalar fields, one can make a decomposition $M = S + i \Phi = S + 2 i \left( \pi^a T^a \right)$
with $\pi^a$ being the pion fields and
$T^a$ being the generators of $SU(2)$ group with the normalization $\mbox{tr} \left( T_a T_b \right) = (1/2) \delta^{ab}$.
With a suitable choice of the Lagrangian of the light meson sector which will not be specialized here, the chiral symmetry can be realized in the Nambu-Goldstone mode. After chiral symmetry breaking, $S$, and therefore $M$ field, acquire vacuum expectation value $\langle M \rangle = f_\pi$ with $f_\pi$ being the pion decay constant. From the Lagrangian \eqref{eq:EffecL}, the $\Delta$ term shifts the masses of the DHBs in the same
direction,
therefore the mass difference between the chiral partners is provided by the $g_\pi$ term as
\begin{eqnarray}
\Delta M_{B;q} =
m_{N_{\textbf{Q},q}^{(\mu)}} - m_{D_{\textbf{Q},q}^{(\mu)}}  =  g_\pi f_\pi .
\label{eq:MassDif}
\end{eqnarray}
The coupling constant $g_\pi$ in the model measures the magnitude of coupling between chiral partners which can be determined as $g_\pi = 4.65$ from the heavy-light spectrum~\cite{Ma:2015lba}. The relation~\eqref{eq:MassDif} between the $f_\pi$ and $g_\pi$ is known as a generalized Goldberger-Treiman relation~\cite{Nowak:1992um,Bardeen:1993ae}.  Then, by using $f_\pi = 92.4$~MeV, the mass
difference for the non-strange doubly heavy baryon is determined as
\begin{eqnarray}
\Delta M_{B;q} = m_{N_{\textbf{Q},q}^{(\mu)}} - m_{D_{\textbf{Q},q}^{(\mu)}} & = & 430~{\rm MeV}.
\label{eq:splitq}
\end{eqnarray}

In terms of the mass difference $\Delta M_{B;q}$, the intermultiplet one-pion transitions of the DHBs in the isospin symmetry limit can be studied. The
relevant partial widths are expressed as
\begin{eqnarray}
\Gamma \left( \Xi_{bc}^{\ast +} \to \Xi_{bc}^{+} + \pi^0 \right) & = & \Gamma \left(
\Xi_{bc}^{\prime + \mu } \to \Xi_{bc}^{+ \mu } + \pi^0 \right) \nonumber\\
& = & \Gamma \left( \Xi_{bc}^{\prime\ast + } \to \Xi_{bc}^{\prime +} + \pi^0 \right) \nonumber\\
& = & \frac{(\Delta M_{B;u,d})^2}{8\pi f_\pi^2} \,\vert p_\pi \vert \ ,
\label{eq:1pionXi}
\end{eqnarray}
where $\vert p_\pi \vert $ is the three-momentum of $\pi$ in the rest frame of the decaying DHB.
The channels including charged pions can be obtained by using the isospin
relation.

To provide some information for the experimental search of the DHBs, we give some explicit results of the masses and strong decay widths of the DHBs. Since the chiral partner structure only yields the mass difference between chiral partners~\eqref{eq:splitq}, we need some ground state masses as reference values. Here, among the existing many calculations (see, e.g., those summarized in Refs.~\cite{Kiselev:2001fw,Kiselev:2002iy,Likhoded:2009zz,Tang:2011fv,Karliner:2014gca}) we choose the results calculated from the nonrelativistic QCD~\cite{Kiselev:1999zj}
in which the heavy diquark is treated in a similar fashion as the present work:
\begin{eqnarray}
m_{\Xi_{bc}} = m_{\Xi_{bc}^\prime} = 6.80 \pm 0.05~{\rm GeV}.
\label{Xi mass}
\end{eqnarray}
Our numerical results are given in Table~\ref{tab:sum}.

\begin{table*}[htb]
\begin{tabular}{c|c|c|c}
\hline
\hline
\qquad Spectrum \qquad\qquad & \qquad Prediction (MeV) \qquad & \qquad Decay channel \qquad\qquad &
\qquad Partial width (MeV) \qquad \\
\hline
$m_{\Xi_{bc}^{\ast}}$ & $ 7230\pm50 $ & $\Xi_{bc}^{\ast +} \to \Xi_{bc}^{+} + \pi^0$ & $ 340$  \\
\hline
$m_{\Xi_{bc}^\mu}$ & $ 6860\pm50\pm20 $ & $ \Xi_{bc}^{\ast +} \to \Xi_{bc}^{0} + \pi^+$ & $ 680$ \\
\hline
$m_{\Xi_{bc}^{\prime \mu}}$ & $ 7290\pm50\pm20 $ & $ \Xi_{bc}^{\prime + \mu} \to \Xi_{bc}^{+ \mu} + \pi^0$ & $
340 $ \\
\hline
$m_{\Xi_{bc}^{\prime \ast}}$ & $ 7230\pm50 $ & $ \Xi_{bc}^{\prime + \mu} \to \Xi_{bc}^{0 \mu} + \pi^+$ & $
680 $ \\
\hline
$m_{\Omega_{bc}^{\ast}}$ & $ 7240\pm70 $ & $ \Xi_{bc}^{\prime\ast +} \to \Xi_{bc}^{\prime +} + \pi^0$ & $ 340 $ \\
\hline
$m_{\Omega_{bc}^{\mu}}$ & $ 6950\pm70\pm20 $ & $ \Omega_{bc}^{\ast} \to \Omega_{bc} + \pi^0 $ & $ 18 \times 10^{-3} $ \\
\hline
$m_{\Omega_{bc}^{\prime \mu}}$ & $ 7300\pm70\pm20 $ & $ \Omega_{bc}^{\prime \mu} \to \Omega_{bc}^{ \mu} +
\pi^0$ & $
20 \times 10^{-3} $ \\
\hline
$m_{\Omega_{bc}^{\prime \ast}}$ & $ 7240\pm70 $ & $ \Omega_{bc}^{\prime \ast} \to \Omega_{bc}^{\prime} +
\pi^0$ & $
18 \times 10^{-3} $ \\
\hline
\hline
\end{tabular}
\caption{\label{tab:sum} Spectrum of the doubly heavy baryons with different heavy quarks and the partial widths of one-pion
intermultiplet transitions. Here, we take $m_{\Xi_{bc}}=m_{\Xi_{bc}^\prime} = 6800\pm 50$~MeV~\cite{Kiselev:1999zj} and $m_{\Omega_{bc}}=m_{\Omega_{bc}^\prime} = 6890\pm70$~MeV~\cite{Kiselev:2000jb}, and $m_{\pi^\pm} = m_{\pi^0} \simeq 140$~MeV as
input. The partial widths are obtained by using the central values of the baryon masses. Other partial widths of intermultiplet transitions can be obtained using the isospin
relation. The first uncertainty is from the reference values in
Eqs.~(\ref{Xi mass}) and (\ref{Omega mass}), and  the second one is from Eq.~\eqref{eq:massdiffinter}.
}
\end{table*}

For the spectroscopy of the DHBs including a strange quark,
by using the spectrum of the
heavy--light meson including a strange quark and also the estimation for the DHBs including the same heavy quarks~\cite{Ma:2015lba}, we predict~\cite{Ma:2015lba,Bardeen:2003kt,Nowak:2003ra,Nowak:2004jg}
\begin{eqnarray}
m_{N_{\textbf{Q},s}^{(\mu)}} - m_{D_{\textbf{Q},s}^{(\mu)}} & = & m_{G_{s}} - m_{H_{s}} = 350~{\rm MeV}.
\label{eq:splits}
\end{eqnarray}
Concerning this magnitude of the mass splitting, one concludes that the dominant decay channel of $\Omega_{bc}^{\ast}$ is not the isospin conserving process $\Omega_{bc}^{\ast} \to \Omega_{bc} + \eta$ but the isospin violating process $\Omega_{bc}^{\ast}
\to \Omega_{bc} + \pi^0$ arising from the $\eta$-$\pi^0$ mixing. The partial decay width is expressed as
\begin{eqnarray}
\Gamma \left( \Omega_{bc}^{\ast } \to \Omega_{bc} + \pi^0 \right) & = & \Gamma \left(
\Omega_{bc}^{\prime\mu} \to \Omega_{bc}^{\mu} + \pi^0 \right) \nonumber\\
& = & \Gamma \left(\Omega_{bc}^{\prime\ast } \to \Omega_{bc}^{\prime} + \pi^0 \right) \nonumber\\
& = & \frac{ (\Delta M_{B;s})^2}{2\pi f_\pi^2} \Delta_{\pi^0\eta}^2\,\vert p_\pi \vert .
\label{eq:1pionOmega}
\end{eqnarray}
where $\Delta_{\pi^0\eta}$ is the magnitude of the $\eta$-$\pi^0$ mixing which was estimated
to be ${} -5.32 \times 10^{-3}$ in Ref.~\cite{Harada:2003kt} based on the two-mixing angle scheme (see, e.g., Ref.~\cite{Harada:1995sj} and references therein).
As the magnitude of the isospin breaking $\eta$-$\pi^0$ mixing is very small, the partial widths in Eq.~\eqref{eq:1pionOmega} are expected to be small which are consistent with the numerical results given in Table.~\ref{tab:sum}.
 The explicit results of the masses and strong decay widths of the Omega DHBs in Table.~\ref{tab:sum} are obtained by chosen the typical reference values~\cite{Kiselev:2000jb}
\begin{eqnarray}
m_{\Omega_{bc}} = m_{\Omega_{bc}^\prime} = 6.89 \pm 0.07~{\rm GeV}.
\label{Omega mass}
\end{eqnarray}
Our numerical results are given in Table~\ref{tab:sum}.

Further, we want to make a comment on the mass splitting of the DHBs in a doublet arising from the heavy quark symmetry breaking effect that is beyond the scope of the present work. The results obtained in
various models or methods listed in Ref.~\cite{Likhoded:2009zz} can be averaged as:
\begin{eqnarray}
m_{\Psi_{bc}^{(\prime)\mu}} - m_{\Psi_{bc}^{(\ast)}} \simeq (60 \pm 20)~{\rm MeV},
\label{eq:massdiffinter}
\end{eqnarray}
which is smaller than the pion mass.
Therefore, the
one-pion transition between DHBs in the same heavy quark multiplet is forbidden for the kinetic reasons. This is dramatically different from the heavy-light meson sector in which, for example, the $D^\ast \to D \pi$ could easily happen.

In summary, the degeneracy of the doubly heavy baryons with different heavy quarks is studied based on the heavy quark symmetry of QCD. We point out that, for the DHBs with the same orbital excitation between heavy diquark and the light constituent, although they can be classified into different heavy quark multiplets, they have the degenerated masses.
For the DHBs with the same quantum number but in different heavy quark multiplets, they do not mix to each other because of the heavy quark spin conservation. In addition, by using the chiral partner structure, we studied the mass splitting and the transition rates between the ground state DHBs and the first excited states of the light quark. The mass splitting is estimated to be about $430$~MeV for the non-strange DHBs and $350$~MeV for the strange DHBs
and, due to kinetic reason, the dominant decay channel of the parity odd strange DHB is an isospin-violating process which, therefore, has a small partial width.

\acknowledgments

Y.-L.~M. is grateful for the hospitality of the Quark-Hadron Theory Group of Nagoya University where part of this work was done.
The work of Y.-L.~M. was supported by the National Science Foundation of China (NSFC) under
Grant No.~11475071 and the Seeds Funding of Jilin University.
M.~H. was supported by the JSPS Grant-in-Aid for Scientific Research (c) No.~24540266.

\end{document}